\documentclass[aps,pra,showpacs,twocolumn]{revtex4-1}

\usepackage[latin1]{inputenc}

\usepackage{graphicx,float}
\usepackage{epsfig}

\usepackage{amsmath}
\usepackage{amssymb}
\usepackage{bm}
\usepackage{bbm}
\usepackage{dsfont}
\usepackage{bbold}
\usepackage{multirow}

\usepackage{hyperref}
\hypersetup{colorlinks=true,linkcolor=blue,citecolor=blue,urlcolor=blue}

\begin{document}

\title{A higher quantum bound for the V\'{e}rtesi-Bene-Bell-inequality and the role of POVMs regarding its threshold detection efficiency}

\date{\today}

\author{J.~F.~Barra}
\author{E.~S.~G\'{o}mez}
\author{G.~Ca\~{n}as}
\author{W.~A.~T.~Nogueira}
\author{L.~Neves}
\author{G.~Lima}
\email{glima@udec.cl}
\affiliation{Departamento de F\'{i}sica, Universidad de Concepci\'{o}n, 160-C Concepci\'{o}n, Chile}
\affiliation{Center for Optics and Photonics, Universidad de Concepci\'{o}n, Chile}
\affiliation{MSI-Nucleus on Advanced Optics, Universidad de Concepci\'{o}n, Chile}



\begin{abstract}
Recently, V\'{e}rtesi and Bene [Phys. Rev. A. {\bf 82}, 062115 (2010)] derived a two-qubit Bell inequality, $I_{CH3}$, which they show to be maximally violated only when more general positive operator valued measures (POVMs) are used instead of the usual von Neumann measurements. Here we consider a general parametrization for the three-element-POVM involved in the Bell test and obtain a higher quantum bound for the $I_{CH3}$-inequality. With a higher quantum bound for $I_{CH3}$, we investigate if there is an experimental setup that can be used for observing that POVMs give higher violations in Bell tests based on this inequality. We analyze the maximum errors supported by the inequality to identify a source of entangled photons that can be used for the test. Then, we study if POVMs are also relevant in the more realistic case that partially entangled states are used in the experiment. Finally, we investigate which are the required efficiencies of the $I_{CH3}$-inequality, and the type of measurements involved, for closing the detection loophole. We obtain that POVMs allow for the lowest threshold detection efficiency, and that it is comparable to the minimal (in the case of two-qubits) required detection efficiency of the Clauser-Horne-Bell-inequality.
\end{abstract}

\maketitle


\section{Introduction}

In quantum mechanics, in general, it is not possible to know in advance the result of a measurement performed on a system. We can only know the spectrum of possible results and the corresponding probabilities \cite{A. Peres}. Quantum measurements are described by a set of operators that act on the Hilbert space of the system under observation. In the case of projective or von Neumann measurements, the operators are orthogonal between them and restricted by the system's dimension. On the other hand, generalized measurements, also known as positive operator valued measures (POVMs), involve measurement operators that only have to be positive, releasing the orthogonality and dimensionality constraints \cite{Nielsen}.

The choice of which type of measurement will be used in certain quantum information protocols, such as remote state preparation \cite{Lo2000, Pati2001, Bennet2001} or quantum state discrimination \cite{Croke}, is important. For example, in the case of unambiguous quantum state discrimination \cite{Ivanovic1987}, Peres \cite{Peres1988} showed that the POVMs found by Dieks \cite{Dieks1988} are actually the ones that minimize the failure probability.

In the Bell inequalities context \cite{Bell}, von Neumann measurements are usually considered \cite{ClauserHorne, CHSH, Weihs, Rowe, Aspect, Torgerson, Kwiat, Rarity, Leach, Lima2010}. Nevertheless, Vértesi and Bene \cite{Vertesi y Bene} recently found that POVMs can also be relevant for Bell tests of quantum nonlocality. They derived a two-qubit Bell inequality, $I_{CH3}$, which involves local measurements with both binary and ternary outcomes. In this Bell inequality, two parties are considered, Alice and Bob, which share copies of an entangled quantum state. Alice can use both von Neumann and POVMs measurements, whereas Bob only uses projective measurements. Furthermore, Alice can choose between three measurement settings $x \in \{0,1,2\}$ and Bob can choose between only two $y \in \{0,1\}$. In this scenario, all the measurements considered have two outcomes, with the exception of the measurement associated to Alice's setting $x=2$, which represents a three-element POVM, that is, a three-outcome measurement. Vértesi and Bene explicitly showed that the $I_{CH3}$ inequality is maximally violated only if POVMs are used instead of the usual von Neumann measurements. However, as they pointed out \cite{Vertesi y Bene}, the maximal violation derived with POVMs corresponds to a lower quantum bound of the inequality, due to the restrictions considered on Alice's and Bob's operators.

In this work we consider a general parametrization for three-element POVMs involved in the inequality, and study further properties of $I_{CH3}$. First, we analyze the inequality's maximal violation. In particular, we restrict ourselves to a maximally entangled state (MES) and maximize numerically $I_{CH3}$ using the Conjugate Gradient (CG) method \cite{metGC}. Due to the fact that we give complete freedom to choose both Alice's and Bob's operators, we obtain (with POVMs) a higher quantum bound for the $I_{CH3}$-inequality with a MES. With a higher quantum bound for $I_{CH3}$ we investigate if there is an experimental setup that can be used for observing that POVMs give higher violations in Bell tests based on this inequality. We analyze the errors supported by the inequality and indicate a specific source of entangled photons that can be used for the test. Then, we study the more realistic case, where partially entangled states (PES) are considered in the experiment. We show that for these states, the maximal violation of $I_{CH3}$ inequality can only be obtained when generalized measurements are also considered.

Finally, to complete our study of $I_{CH3}$, we search for the threshold detection efficiencies required by this inequality, and which are the type of measurements associated to them, for closing the detection loophole. Considering our general parametrization for three-element POVMs, and using the CG method, we perform a minimization of the required detection efficiencies of $I_{CH3}$. We obtain that the lowest required efficiency is given by a POVM and, surprisingly, that its value is comparable to the minimum detection efficiency required by the Clauser-Horne inequality \cite{Eberhard, Larsson2}, which corresponds to the lowest known value in the two-qubit case. Hence, POVMs, in addition of being relevant for the maximal violation of the $I_{CH3}$-inequality, can also be relevant for the effort of performing loophole-free Bell tests.


\section{A Higher quantum bound for the $I_{CH3}$ Bell inequality}

\subsection{Overview of Vértesi and Bene's work}

Vértesi and Bene \cite{Vertesi y Bene} consider a standard Bell scenario \cite{Bell} in which two parties, Alice and Bob, share copies of a quatum state $\rho$. Alice's and Bob's measurements are labeled $x$ and $y$, where their respective outputs are denoted by $a$ and $b$. The operators corresponding to these measurements are $M_a^x$ for Alice and $M_b^y$ for Bob. Hence, the joint conditional probabilities are given by $p(ab|xy)~=tr(\rho M_a^x \otimes M_b^y)$. The authors focus their study on two qubits in a maximally entangled state $\rho = |\phi^+\rangle\langle\phi^+|$, with $|\phi^+\rangle = 1/\sqrt{2}(|00\rangle + |11\rangle)$.

The Bell inequality presented in \cite{Vertesi y Bene}, is composed of the probabilities involved in the $I_{CH}$ inequality \cite{CHSH, ClauserHorne}, and an expression that considers a three-outcome generalized measurement on Alice's side, named $I_3$
\begin{equation}
I_{CH3} \equiv c I_{CH} + I_3 \leq 1, \label{eq:ICH3}
\end{equation} with $c > 0$. What we have then, is that Eq.~(\ref{eq:ICH3}) represents a family of inequalities, where each of the $I_{CH3}$ inequalities are defined when the $c$ value is set. The $c$ parameter also tells us how dominant the $I_{CH}$ inequality probabilities are over the $I_3$ expression. The $I_{CH3}$ inequality is explicitly given by
\begin{eqnarray}
I_{CH3} &= &c\ [p(00|00)+p(00|01)+p(00|10)-p(00|11) \nonumber\\
&&-\ p_A(0|0)-p_B(0|0)]+p(00|20)+p(00|21) \nonumber\\
&&+\ p(10|20)-p(10|21)-p_A(0|2) \nonumber\\
&&-\ (1-1/\sqrt{2})p_A(1|2) \leq 1. \label{eq:ICH3_2}
\end{eqnarray}
From Eq. (\ref{eq:ICH3_2}), we can clearly observe that Alice possesses three measurement configurations $x \in \{0,1,2\}$ and Bob two $y \in \{0,1\}$. All measurements have binary outcomes $a,b \in \{0,1\}$, except Alice's configuration $x=2$ corresponding to a three-outcome measurement $a \in \{0,1,2\}$.

In order to demonstrate that POVMs give a maximal violation of $I_{CH3}$, Vértesi and Bene first calculate the inequality's maximum using only von Neumann measurements. In this case, the configuration $x=2$ corresponds to a projective measurement, different from the ones already used by Alice on $I_{CH}$. Since we find ourselves working with qubit states, $\sum_{a=0}^2$rank$(M_a^{x=2}) = 2$ must be satisfied, entailing that at least one of the three $M_a^{x=2}$ operators will always correspond to the null operator. In this way, there are six possible combinations  between the operator's $M_{a=0}^{x=2}$ and $M_{a=1}^{x=2}$ ranks. According to this, it is possible to derive six new Bell inequalities involving only projective measurements. They are denoted by $I_{00}$, $I_{01}$, $I_{10}$, $I_{11}$, $I_{02}$ and $I_{20}$ depending on the rank of the projectors $M_{a=0}^{x=2}$ and $M_{a=1}^{x=2}$ \cite{Vertesi y Bene}.

The quantum maximum of these six inequalities are calculated considering that Alice's and Bob's projectors are parametrized by a unit vector $\vec{v} = (v_x,v_y,v_z)$
\begin{eqnarray}
M_{i=0}^m (\vec{v}) = \frac{1}{2}(\mathds{1} + \vec{v}\cdot\vec{\sigma}), \\
M_{i=1}^m (\vec{v}) = \frac{1}{2}(\mathds{1} - \vec{v}\cdot\vec{\sigma}),
\end{eqnarray} with $\vec{\sigma} = (\sigma_x,\sigma_y,\sigma_z)$, $m=x,y$, and $i=a,b$. In this way, using projective measurements, the authors obtain that the $I_{CH3}$ quantum maximum for $c > 0$ corresponds to the $I_{10}$ inequality's maximum. It is given by
\begin{equation}
 	\mathop{\mbox{max}} \limits_{proj,\phi^+}I_{CH3}=\frac{-c+\sqrt{c^2+(c+1)^2}}{2}. \label{eq:maxI10}
\end{equation}

To obtain the maximum of the $I_{CH3}$-inequality using POVMs, certain restrictions are imposed upon the measurement operators \cite{Vertesi y Bene}. They use POVMs whose elements are proportional to rank-1 projectors with real coefficients, and set Bob's projectors (and consequently Alice's projectors associated to $I_{CH}$) in the orientations that maximizes the $I_{CH}$-inequality. For the state $|\phi^+\rangle$, the $I_{CH}$ maximum obtained is given by $(\sqrt{2}-1)/2$, whereas for the expression $I_3$, the value 0.3788 is obtained. Alice's POVM that gives this value is explicitly shown in \cite{Vertesi y Bene}. The sum of these values provides a quantum bound for $I_{CH3}$ with POVMs on state $|\phi^+\rangle$. It is given by
\begin{equation}
  \mathop{\mbox{max}} \limits_{POVM,\phi^+} I_{CH3} \geq \frac{c(\sqrt{2}-1)}{2} + 0.3788, \label{eq:maxPOVM}
\end{equation} which is always higher than the maximum obtained with projective measurements when $c \geq 3$. Nevertheless, the result of Eq. (\ref{eq:maxPOVM}) can be viewed as a lower quantum bound for $I_{CH3}$ inequality, due the restrictions imposed on Alice and Bob's measurement operators \cite{Vertesi y Bene}.


\subsection{Maximum of $I_{CH3}$-inequality with a MES}

In our study we perform a global maximization of $I_{CH3}$, allowing complete freedom in the choices of Alice's and Bob's operators. We  consider a pair of qubits in the MES and that measurement operators  ($M_i^m$), corresponding to projective measurements, are parametrized according to the following orthonormal base
\begin{eqnarray}
|v\rangle _{{\phi}_k}&=&\sin\phi_k|+\rangle+e^{iv_{\phi_k}}\cos\phi_k|-\rangle, \label{eq:PROJv}\\
|u\rangle _{{\phi}_k}&=&\cos\phi_k|+\rangle-e^{iv_{\phi_k}}\sin\phi_k|-\rangle, \label{eq:PROJu}
\end{eqnarray}
where vectors $|v\rangle _{\phi_k}$ and $|u\rangle _{{\phi}_k}$ are associated to the outcomes $a,b=0$ and $a,b=1$ respectively. The angle $\phi_k$ represents the measurement orientation/configuration. The states $|\pm\rangle$ define the logical base. Table \ref{tb:1} shows the correspondence of index $k$ with the measurements settings in the Vértesi and Bene's notation.

\begin{table}[th]
\centering
\begin{tabular}{c|c|c}
Alice & $x=0$ & $k=1$ \\
& $x=1$ & $k=2$  \\ \hline
Bob & $y=0$ & $k=3$ \\
& $y=1$ & $k=4$
\end{tabular}
\caption{Correspondence between the different notations for the measurement configurations discussed in the main text.}
\label{tb:1}
\end{table}

The generalized measurement corresponding to Alice's configuration $x=2$ is parametrized by a general three-element POVM. Just like in \cite{Vertesi y Bene}, its elements are denoted by $M_i$ ($i=a$). Our POVMs elements are given by
\begin{small}
\begin{eqnarray}
M_0 &=& \left(
\begin{array}{cc}
	\cos^2\theta\cos^2\varphi & -e^{i\omega_0}\cos^2\eta\cos^2\gamma\\
	-e^{-i\omega_0}\cos^2\eta\cos^2\gamma & \cos^2\chi\cos^2\mu
\end{array}
\right), \label{M0}\\
M_1 &=&\left(
\begin{array}{cc}
	\sin^2\theta & e^{i\omega_1}\cos^2\eta\\
	e^{-i\omega_1}\cos^2\eta & \sin^2\chi
\end{array}
\right). \label{M1}
\end{eqnarray}
\end{small}By definition, $\sum_i M_i = \mathds{1}$ must be satisfied, hence, $M_2 =~ \mathds{1} - M_0 - M_1$. To satisfy the positivity condition of the measurement operators, the following inequalities must be fulfilled during the maximization
\begin{small}
\begin{eqnarray}
\cos^2\theta\cos^2\varphi\cos^2\chi\cos^2\mu &\geq& \cos^4\eta\cos^4\gamma, \label{cond1}\\
\sin^2\theta\sin^2\chi &\geq& \cos^4\eta, \label{cond2}\\
\cos^2\theta\sin^2\varphi\cos^2\chi\sin^2\mu &\geq& \cos^4\gamma(e^{i\omega_0}\cos^2\gamma-e^{i\omega_1}) \nonumber\\
&& \times(e^{-i\omega_0}\cos^2\gamma-e^{-i\omega_1}) \label{cond3} .
\end{eqnarray}
\end{small}

The maximum of the $I_{CH3}$-inequality can be obtained with an exhaustive search done with the well-known conjugated gradient numerical method \cite{metGC}. Considering our parametrization for the measurement operators, the $I_{CH3}$ inequality can be seen as a $16$ variable-function. These 16 variables (shown in Table \ref{tb:2}) define the parameter space in which the maximization is performed. The CG method uses the local gradient in a given initial point of the parameter space to reach the nearest local maximum point. The search of the maximum is done with the numeric precision set in $10^{-6}$. In order to map all the local maximal and decide which is the global maximum, we ran the CG program for a large sample of initial points in the parameter space. We considered samples of size greater than $10^4$ points for each simulation. To verify that the maximum reached was the global maximum, we reiterated the search.

\begin{table}[th]
\centering
\begin{tabular}{c|c|c}
& $I_{CH}$ & $I_{3}$ \\ \hline
\multirow{2}{1.5cm}{Alice} & $\phi_1$, $\nu_{\phi_1}$ & $\theta$, $\varphi$, $\eta$, $\gamma$ \\
			& $\phi_2$, $\nu_{\phi_2}$ & $\chi$, $\mu$, $\omega_0$, $\omega_1$ \\ \hline
\multirow{2}{1.5cm}{Bob} 	& \multicolumn{2}{c}{$\phi_3$, $\nu_{\phi_3}$}  \\
			&	\multicolumn{2}{c}{$\phi_4$, $\nu_{\phi_4}$} 	
\end{tabular}
\caption{$I_{CH3}$ inequality variables in accordance with the parties and the $I_{CH}$ and $I_3$ expressions.}
\label{tb:2}
\end{table}

During the maximization of the $I_{CH3}$-inequality, using the CG method, we didn't constraint nor Alice's or Bob's operators. However, when searching for the global maximum, it is necessary to include the positivity conditions (\ref{cond1}), (\ref{cond2}) and (\ref{cond3}) for Alice's POVMs. It is important to mention that our program does not force Alice's $x=2$ measurement operator to be a three-element POVM. That is, if there is a configuration where the maximum is obtained with projective measurements, the program gives them as the maximization process result. Therefore, from the numerical search, we obtain the $I_{CH3}$ maximum and the type of measurement associated to it.

\begin{figure}[th]
	\centering
		\includegraphics[width=0.35\textwidth,angle=-90]{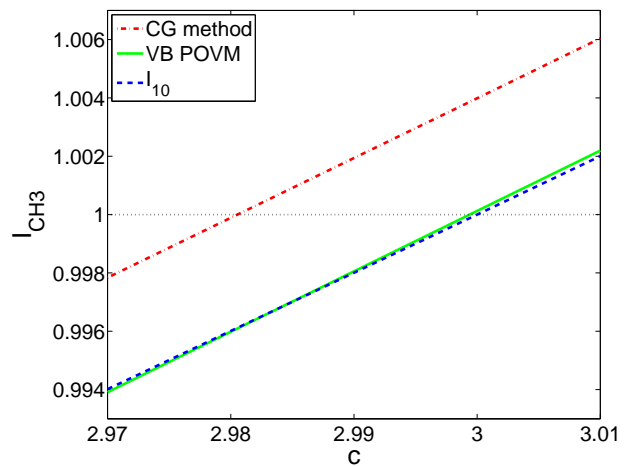}  
	\caption{(Color online) Plot of the $I_{CH3}$ inequality for a MES. The red (dashed-dotted line) curve corresponds to the maximum obtained with the CG method, without restrictions on both Alice and Bob's measurement operators. The green (solid line) curve is the Vértesi and Bene's maximum \cite{Vertesi y Bene}. The blue (dashed line) curve corresponds to the $I_{10}$ maximum, which denotes the $I_{CH3}$ maximum using only projective measurement.}
	\label{fig:MES_POVM_GenvsVB}
\end{figure}

\begin{figure}[th]
	\centering
		\includegraphics[width=0.35\textwidth,angle=-90]{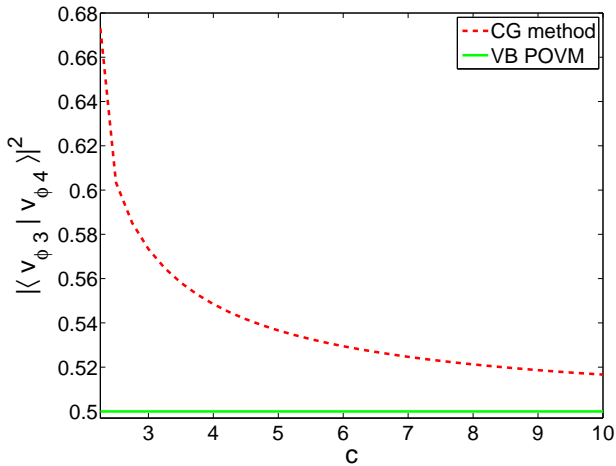}       
	\caption{(Color online) Inner product between the states related with Bob's measurement projectors. The red (dashed line) curve is obtained according to the $I_{CH3}$ maximization with the CG method. The green (solid line) curve corresponds to the inner product between the states related with the measurement projectors that maximize $I_{CH}$.}
	\label{fig:MES_PRODINT}
\end{figure}

Fig. \ref{fig:MES_POVM_GenvsVB} shows the maximum for the MES, obtained by our $I_{CH3}$ maximization, as a function of the parameter $c$. We observe that $I_{CH3}$ maxima obtained are higher than the Vértesi and Bene's maxima given in Eq. (\ref{eq:maxPOVM}). In our case, maximal violations of $I_{CH3}$ were always obtained with three-element POVMs. For each $c$ value, we have a different POVM associated to the maximal violations of $I_{CH3}$. For example, in the case $c=3$ we have $I_{CH3}$ violation with value 1.004, and the corresponding POVM is
\begin{small}
\begin{eqnarray}
M_0 &=& \left(
\begin{array}{cc}
	0.8890 & -0.0818 + 0.1047i\\
	-0.0818 - 0.1047i & 0.0198
\end{array}
\right), \label{M01}\\
M_1 &=&\left(
\begin{array}{cc}
	0.0553 & -0.1023 - 0.0995i\\
	-0.1023 + 0.0995i & 0.3680
\end{array}
\right), \label{M11}
\end{eqnarray}
\end{small}with $M_2 = \mathds{1} - M_0 - M_1$. Our POVMs elements are defined with numeric precision $10^{-6}$, but they are shown above with less digits to make it clearer. The POVMs corresponding to $I_{CH3}$ maximum are always of rank-1 (within our numeric precision) as expected when dealing with the maximization of a Bell inequality \cite{Gisin2008}.

One of the reasons why Vértesi and Bene's result is considered a lower bound (see Ref. \cite{Vertesi y Bene}) is that, as previously mentioned, they set Bob's projectors in the orientations that maximize $I_{CH}$. In this case, the states related with such measurement projectors have an inner product $|\langle v_{\phi_3}|v_{\phi_4}\rangle|^2 = 0.5$. In our case, Bob's projectors are free, and one can observe in Fig.~\ref{fig:MES_PRODINT} how this inner product varies as a function of $c$. One can see that when the $c$ value is small, the optimal orientations for Bob's projectors in $I_{CH3}$ are far from those that maximize $I_{CH}$. However, as $c$ increases, these orientations come asymptotically closer to the optimal measurements of $I_{CH}$.

\begin{figure}[ht]
	\centering
		\includegraphics[width=0.35\textwidth,angle=-90]{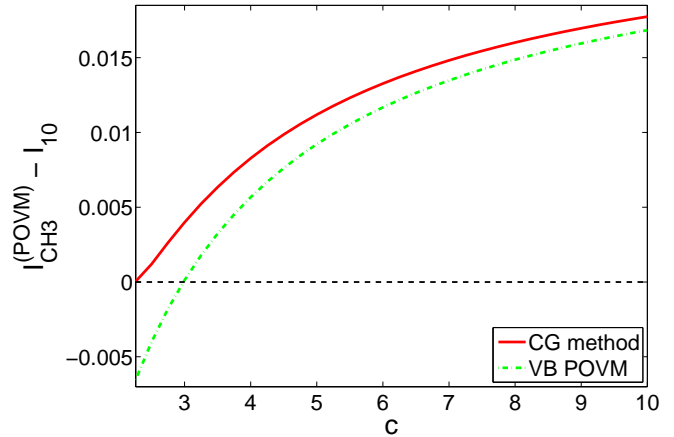}   
	\caption{(Color online) Difference between the maxima of $I_{CH3}$ with POVMs and $I_{10}$ [Eq.~\ref{eq:maxI10}] for a MES. The red (solid line) curve corresponds to the difference between $I_{CH3}$ maxima obtained with the CG method and $I_{10}$. The green (dashed-dotted line) curve is the difference between the $I_{CH3}$ maxima with Vértesi and Bene's POVM and $I_{10}$.}
	\label{fig:DIF_MES_POVMGen_POVMVB}
\end{figure}

\subsection{Maximum error supported for the observation of POVMs relevance}

Another way to observe the POVMs relevance in the violation of $I_{CH3}$-inequality (that is, that POVMs give higher violations of $I_{CH3}$ than von Neumann measurements) is through the difference between the maxima obtained with POVMs and von Neumann measurements, as shown in Fig. \ref{fig:DIF_MES_POVMGen_POVMVB}. Although we obtain positive values for this difference, these values are too small. In order to determine the possibility of observing experimentally that POVMs give higher violations for $I_{CH3}$ Bell inequality, we study this difference when typical experimental errors are taken into account. Generally, when noise effects are considered in a certain Bell inequality they are taken to be of a particular form, like white noise \cite{Vertesi y Bene,Werner1989,Brunner2007} or colored noise \cite{Cabello2005}, and added to the initial state. In our case, we take a more practical approach and study how much can be the reduction in the $I_{CH3}$-inequality violation (due to experimental errors) to observe the POVMs relevance. In doing so, we can compare our results directly with the errors shown on experimental works \cite{Weihs, Rowe, Aspect, Torgerson, Kwiat, Rarity, Leach, Lima2010}.

The $I_{CH3}$ error is given by
\begin{equation}
\Delta I_{CH3} = c \Delta I_{CH} + \Delta I_3,
\end{equation} where $\Delta I_{CH}$ is the error associated to the $I_{CH}$ inequality and $\Delta I_3$ is the error associated to the expression $I_3$.

\begin{figure}[ht]
	\centering
		\includegraphics[width=0.35\textwidth,angle=-90]{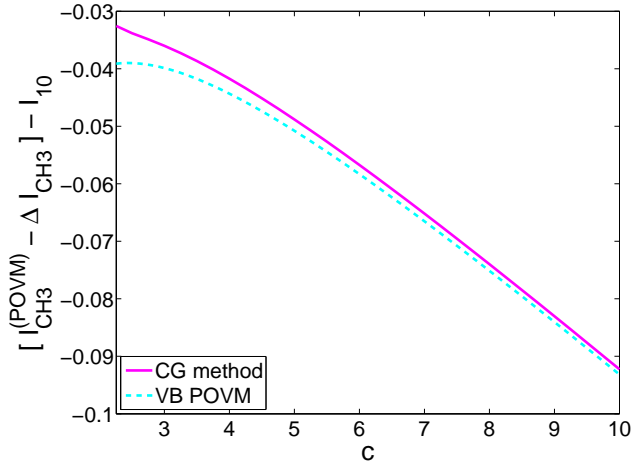}  
	\caption{(Color online) Difference between $I_{CH3}$ with POVMs and $I_{10}$ for a MES, while considering $\Delta I_{CH} = \Delta I_3 = 0.01$. The pink (solid line) curve corresponds to the difference between $I_{CH3}$ and $I_{10}$ maxima obtained with CG method. The cyan (dashed line) curve is the difference between $I_{CH3}$ maxima with Vértesi and Bene's POVM and $I_{10}$.}
	\label{fig:DIF_MES_ERROR_GenVBPOVM}
\end{figure}

In an experiment, POVMs relevance is proven when
\begin{equation}
I_{CH3}^{(POVM)} - \Delta I_{CH3} > I_{CH3}^{(PROJ)}.
\end{equation} Let's recall that for a MES, the $I_{10}$ inequality is the one that renders the maxima of $I_{CH3}$ with projective measurements (i.e., $I_{CH3}^{(PROJ)} = I_{10}$) \cite{Vertesi y Bene}. When the error is considered, there also has to be a violation of the inequality. Therefore,
\begin{equation}
I_{CH3}^{(POVM)} - \Delta I_{CH3} > 1. \label{eq:viol}
\end{equation} In our case, we consider symmetrical errors $\Delta I_{CH} = \Delta I_3$, since the number of probabilities considered in $I_{CH}$ and $I_3$ are the same. In Fig. \ref{fig:DIF_MES_ERROR_GenVBPOVM} we can observe how the difference behaves when $\Delta I_{CH} = \Delta I_3 = 0.01$, corresponding to the standard error of photonic Bell experiments \cite{Torgerson, Kwiat, Rarity, Leach, Lima2010}. One can see that an experiment designed to demonstrate POVMs relevance for the violation of the $I_{CH3}$ inequality does not support this error level, since we only obtain negative values in the differences.

\begin{figure}[ht]
	\centering
		\includegraphics[width=0.35\textwidth,angle=-90]{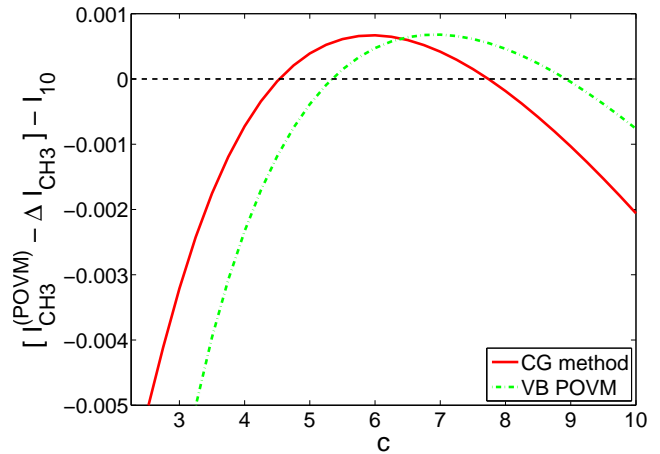}  
	\caption{(Color online) Difference between $I_{CH3}$ with POVMs and $I_{10}$ considering the highest supported error $\Delta I_{CH3}$ for a MES. The red (solid line) curve corresponds to the difference between the $I_{CH3}$ maxima obtained by the CG method and $I_{10}$ maximum using $\Delta I_{CH} = \Delta I_3 = 0.0018$. The green (dashed-dotted line) curve corresponds to the difference between the $I_{CH3}$ maxima with the Vértesi and Bene's POVM and $I_{10}$ using $\Delta I_{CH} = \Delta I_3 = 0.0016$.}
	\label{fig:MES_ERR_OP}
\end{figure}

Because the relevance of POVMs into the $I_{CH3}$ violation cannot be observed while considering standard errors of photonic Bell experiments, we searched for the error tolerance. In doing so, we lowered the error value considered in $I_{CH}$ and $I_3$, until we reached positive value differences. As expected, Vértesi and Bene's POVM has less tolerance to the error than our POVMs, being the highest error supported by the first $\Delta I_{CH} = \Delta I_3 = 0.0016$. For our POVMs, obtained with the CG method, the highest error supported corresponds to $\Delta I_{CH} = \Delta I_3 = 0.0018$. In Fig.~\ref{fig:MES_ERR_OP} we show the differences $[I_{CH3}^{(POVM)}-~\Delta I_{CH3}]-~I_{10}$ curves when these errors are considered.

From Fig.~\ref{fig:MES_ERR_OP} one can see that the highest value of the difference between the maximum of $I_{CH3}$ and $I_{10}$, obtained with the CG method, and considering the maximum supported error, occurs at $c=6$ and its value is $[I_{CH3}^{(POVM)}-~\Delta I_{CH3}]-~ I_{10} = 0.000668$. For this $c$ value we have $I_{CH3}^{(POVM)} - \Delta I_{CH3} = 1.610440$, so the condition (\ref{eq:viol}) is fulfilled. The corresponding POVM to $c=6$, i.e., the optimal POVM possesses the following elements
\begin{small}
\begin{eqnarray}
M_0 &=& \left(
\begin{array}{cc}
	0.8069 & -0.1876 + 0.1650i\\
	-0.1876 - 0.1650i & 0.0774
\end{array}
\right), \label{M02}\\
M_1 &=&\left(
\begin{array}{cc}
	0.1730 & -0.1019 - 0.2281i\\
	-0.1019 + 0.2281i & 0.3608
\end{array}
\right), \label{M12}
\end{eqnarray}
\end{small}with $M_2 = \mathds{1} - M_0 - M_1$.

The error supported by this measurement setting [Eq.~(\ref{M02}) and Eq.~(\ref{M12})], to demonstrate the relevance of POVMs over the projective measurements on the violation of the $I_{CH3}$ inequality, is one order of magnitude below the standard errors of photonic Bell experiments \cite{Torgerson, Kwiat, Rarity, Leach, Lima2010}. However, there are recent experimental setups, based on a twin-photon ultra-bright source, that have reported violations of Bell inequalities with precisions of the order $\Delta = 0.001$ \cite{Shapiro2004,Kim2006,Smith2012}. Therefore, this new kind of source can be used, in principle, for an experimental demonstration of POVMs relevance in quantum nonlocality tests.

\section{Maximum of $I_{CH3}$ with partially entangled states}

\begin{figure}[ht]
	\centering
		\includegraphics[width=0.35\textwidth,angle=-90]{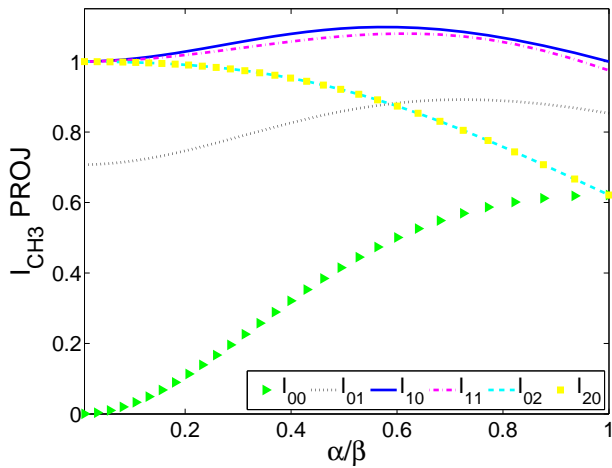}  
	\caption{(Color online) $I_{00}$, $I_{01}$, $I_{10}$, $I_{11}$, $I_{02}$ and $I_{20}$ inequalities maxima considering the entire range of $\alpha/\beta$ for $c=3$.}
	\label{fig:PROJ_c3}
\end{figure}

Here we study the maximal violation of $I_{CH3}$ using PES. Once again, we use the CG method to carry out the $I_{CH3}$ maximization for different $c$ values and considering various degrees of entanglement for the two-qubit state. We proceed in the same way as for MES, searching first for the maxima with projective measurements, and secondly for the maxima using POVMs. Then, we compare them through the differences $I_{CH3}^{(POVM)} - I_{CH3}^{(PROJ)}$ and investigate if the POVMs are also relevant for the violation of $I_{CH3}$ when partially entangled states are considered. Finally, we investigate which are the optimal states to observe the POVMs relevance in this Bell test. This is done by comparing the differences $I_{CH3}^{(POVM)} - I_{CH3}^{(PROJ)}$ for states with different degrees of entanglement.


\subsection{Two-qubit state parametrization}

We consider two-qubit pure states in the Schmidt decomposition, in the form $|\Psi\rangle =\alpha|01\rangle +\beta |10\rangle$, where coefficients $\alpha$ and $\beta$ are real and positive. In this case, the concurrence $C(\Psi)$ is given by $C(\Psi)=2\alpha\beta$ \cite{Wootters2001}. The different degrees of entanglement will be represented through the parameter $\alpha/\beta$, and its relation with the concurrence is
\begin{equation}
\frac{\alpha}{\beta} = \frac{C(\Psi)}{2\beta^2}. \label{eq:ab}
\end{equation}

Thus, the states with $\alpha/\beta = 1$ correspond to maximally entangled states. Those with values $0<\alpha/\beta<1$ are partially entangled states and if $\alpha/\beta = 0$, product states are obtained. This parameter has been extensively used in Bell inequalities studies \cite{Hardy93,Hardy97}.

\subsection{Maxima with projective measurements}

\begin{figure}[ht]	
	\includegraphics[width=0.35\textwidth,angle=-90]{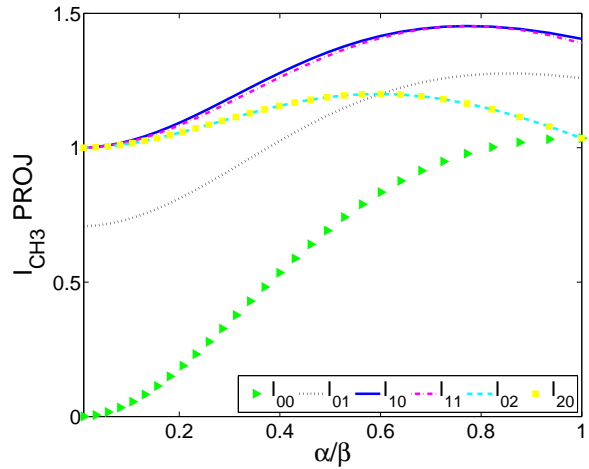}   
	\caption{(Color online) $I_{00}$, $I_{01}$, $I_{10}$, $I_{11}$, $I_{02}$ and $I_{20}$ inequalities maxima considering the entire range of values for $\alpha/\beta$ and $c=5$.}
	\label{fig:PROJ_c5}
\end{figure}

\begin{figure}[ht]	
	\includegraphics[width=0.35\textwidth,angle=-90]{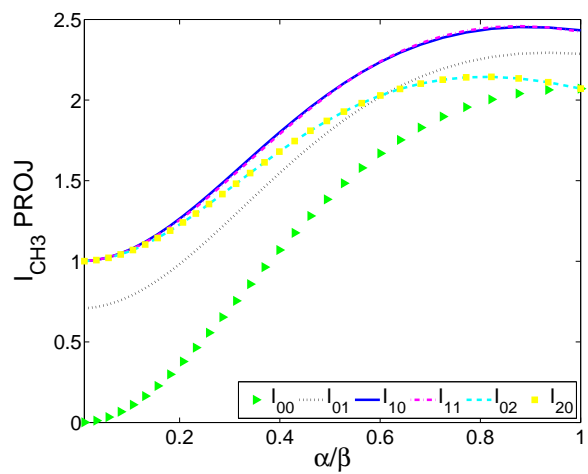}   
	\caption{(Color online) $I_{00}$, $I_{01}$, $I_{10}$, $I_{11}$, $I_{02}$ and $I_{20}$ inequalities maxima considering the entire range of values for $\alpha/\beta$ and $c=10$.}
	\label{fig:PROJ_c10}
\end{figure}

We previously showed that for a MES the $I_{CH3}$ maximum with projective measurements is given by the maximum of the $I_{10}$ inequality. Now we have to find the maxima of the inequalities $I_{00}$, $I_{01}$, $I_{10}$, $I_{11}$, $I_{02}$ and $I_{20}$, considering all the value range of $\alpha/\beta$, and different $c$ values. This is how we determined the $I_{CH3}$ maximum with projective measurements when $0<\alpha/\beta<1$. The expressions for these inequalities are explicitly given in \cite{Vertesi y Bene}. To maximize these inequalities, we used again the CG method. In the case of a MES our results coincide with the ones obtained by Vértesi and Bene.

The curves of these inequalities' maxima for $c=3$ are plotted in Fig. \ref{fig:PROJ_c3}. From these curves, the inequality $I_{10}$ can be identified as dominant over the entire range of $\alpha/\beta$. However, in Fig.~\ref{fig:PROJ_c5} we can notice that when the $c$ value is slightly increased to $c=5$, the inequality $I_{11}$ also starts to be relevant for a certain range of states. Furthermore, as shown in Fig.~\ref{fig:PROJ_c10}, as the $c$ value increases the inequalities $I_{10}$ and $I_{11}$ continue being dominant. It can also be seen that the range of states in which $I_{11}$ is relevant, increases too. It is possible to show that the only relevant inequalities are $I_{10}$ and $I_{11}$ through all $\alpha/\beta$ for any value of $c$.

\subsection{Maxima with POVMs}

We performed the maximization of $I_{CH3}$ with the CG method, as previously discussed, for states with different degrees of entanglement and distinct $c$ values.

\begin{figure}[ht]
	\centering
		\includegraphics[width=0.35\textwidth,angle=-90]{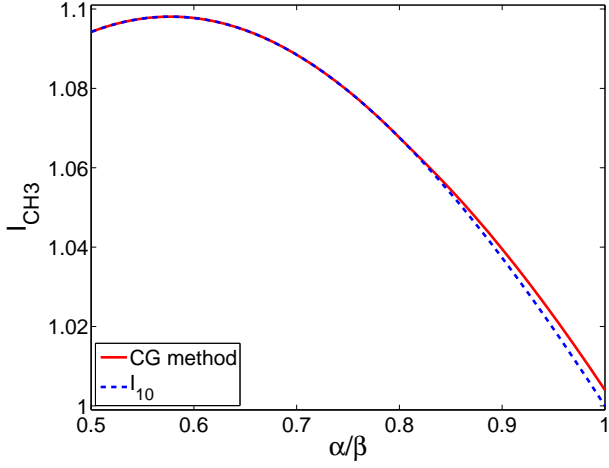}  
	\caption{(Color online) $I_{CH3}$ maxima as a function of $\alpha/\beta$ for $c=3$. The red (solid line) curve represents $I_{CH3}$ maxima using the CG method. The blue (dashed line) curve corresponds to the $I_{CH3}$ maxima with projective measurements.}
	\label{fig:ICH3_I10_ab05_MES_c3}
\end{figure}

Fig. \ref{fig:ICH3_I10_ab05_MES_c3} displays the behavior of $I_{CH3}$, as a function of $\alpha/\beta$, for $c=3$. One can see that there are PES for which the maxima given by POVMs exceed the maxima of $I_{10}$. It can also be observed that for this $c$ value, the maxima violation of $I_{CH3}$ is given for a partially entangled state rather than for the maximally entangled state, either with POVMs or with projective measurements. This is another example that nonlocality and entanglement are not the same physical concept \cite{Vidick2011}. For $c=3$ and $\alpha/\beta=~0.9006$, the measurement that gives the maximum of $I_{CH3}$ is a POVM whose elements are
\begin{small}
\begin{eqnarray}
M_0 &=& \left(
\begin{array}{cc}
0.0365 & -0.0640 - 0.1687i\\
-0.0640 + 0.1687i & 0.8933
\end{array}
\right), \label{M03}\\
M_1 &=&\left(
\begin{array}{cc}
0.3140 & 0.0649 + 0.1712i\\
0.0649 - 0.1712i & 0.1067
\end{array}
\right), \label{M13}
\end{eqnarray}
\end{small}where $M_2 = \mathds{1} - M_0 - M_1$.

\begin{figure}[htbp]
	\centering
		\includegraphics[width=0.35\textwidth,angle=-90]{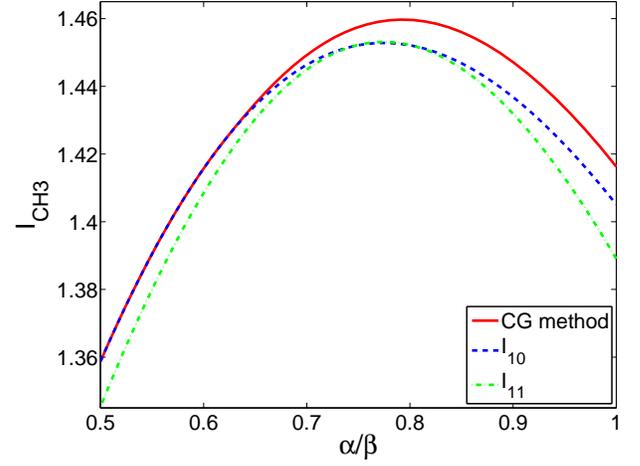}  
	\caption{(Color online) $I_{CH3}$ maxima as a function of $\alpha/\beta$ for $c=5$. The red (solid line) curve represents $I_{CH3}$ maxima using the CG method as described in the main text. The blue (dashed line) and the green (dashed-dotted line) curves correspond to the $I_{CH3}$ maxima with projective measurements, the former is the $I_{10}$ (blue) and the latter $I_{11}$ (green).}
	\label{fig:ICH3_I10_I11_ab05_MES_c5}
\end{figure}

\begin{figure}[htbp]
	\centering
		\includegraphics[width=0.35\textwidth,angle=-90]{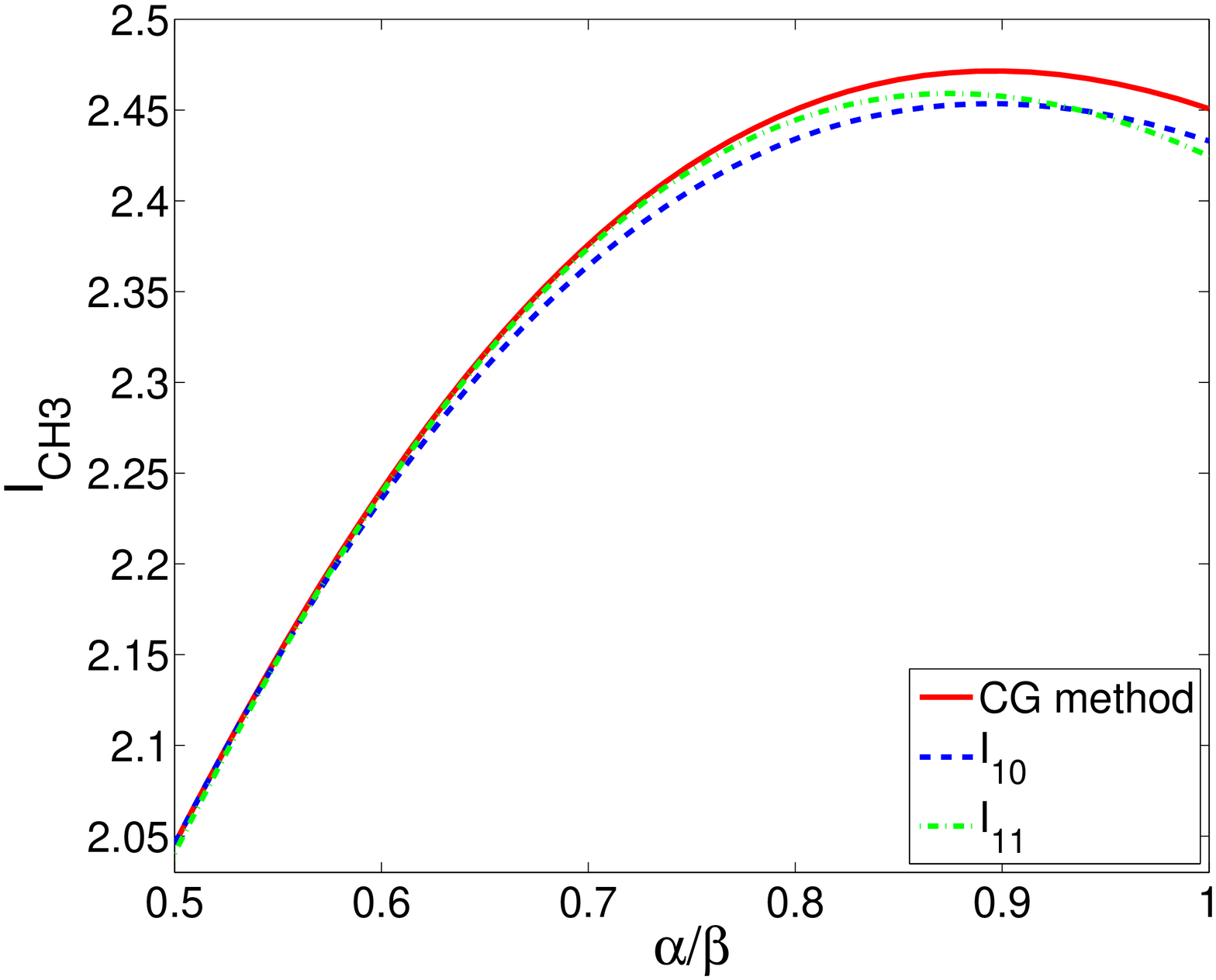}   
	\caption{(Color online) $I_{CH3}$ maxima in fuction of $\alpha/\beta$ for $c=10$. The red (solid line) curve represents $I_{CH3}$ maxima using the CG method as described in the main text. The blue (dashed line) and the green (dashed-dotted line) curves correspond to the $I_{CH3}$ maxima with projective measurements, the former is the $I_{10}$ (blue) and the latter $I_{11}$ (green).}
	\label{fig:ICH3_I10_I11_ab05_MES_c10}
\end{figure}

In Fig. \ref{fig:ICH3_I10_I11_ab05_MES_c5} we present the results for $c=5$. Again we have that the maximal violation of $I_{CH3}$ doesn't correspond to a MES. In this case, there are also PES for which POVMs render the maximum of $I_{CH3}$ above the maximum of $I_{10}$ and $I_{11}$. In fact, the set of PES for which POVMs are the optimal measurements for the violation of $I_{CH3}$ increases as the $c$ value increases. The POVM that maximizes $I_{CH3}$ in $\alpha/\beta = 0.8025$ is given by the following elements
\begin{small}
\begin{eqnarray}
M_0 &=& \left(
\begin{array}{cc}
	0.0756 & 0.1302 - 0.2187i\\
  0.1302 + 0.2187i & 0.8566
\end{array}
\right), \label{M04}\\
M_1 &=&\left(
\begin{array}{cc}
	0.4669 & -0.1323 + 0.2223i\\
	-0.1323 - 0.2223i & 0.1434
\end{array}
\right), \label{M14}
\end{eqnarray}
\end{small}with $M_2 = \mathds{1} - M_0 - M_1$.

Finally, Fig.~\ref{fig:ICH3_I10_I11_ab05_MES_c10} gives the results obtained for $c=10$. By the form of the inequality $I_{CH3}$ [Eq.~(\ref{eq:ICH3})], we have that as the $c$ value increases, the inequality $I_{CH}$ becomes dominant over $I_3$. This can be seen in the $I_{CH3}$ curve, since it starts to resemble the $I_{CH}$ curve. The $I_{CH3}$ maximum for $\alpha/\beta = 0.7067$ is given by a POVM which possesses the following elements
\begin{small}
\begin{eqnarray}
M_0 &=& \left(
\begin{array}{cc}
	0.1033 & 0.2861 + 0.0789i\\
  0.2861 - 0.0789i & 0.8528
\end{array}
\right), \label{M05}\\
M_1 &=&\left(
\begin{array}{cc}
	0.6147 & -0.2899 - 0.0799i\\
	-0.2899 + 0.0799i & 0.1471
\end{array}
\right), \label{M15}
\end{eqnarray}
\end{small}where $M_2 = \mathds{1} - M_0 - M_1$.

From Figs. \ref{fig:ICH3_I10_ab05_MES_c3}-\ref{fig:ICH3_I10_I11_ab05_MES_c10} we can conclude that, even though the $I_{CH}$ inequality becomes dominant as the parameter $c$ increases, there is an increase in the set/range of PES for which POVMs are relevant.

\begin{figure}[ht]
	\centering
		\includegraphics[width=0.35\textwidth,angle=-90]{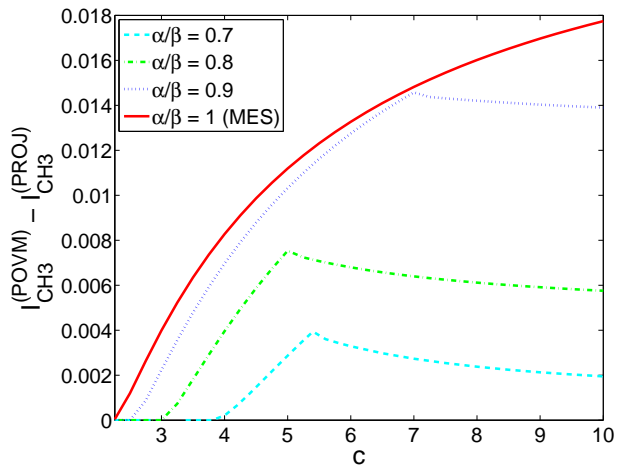}  
		\caption{(Color online) Difference between $I_{CH3}$ with POVMs and $I_{CH3}$ with projective measurements. The red (solid line) curve corresponds to $\alpha/\beta=1$, which is a continuous curve because there is only one relevant projective based Bell inequality, $I_{10}$. The blue (dotted line) curve correspond to $\alpha/\beta=0.9$, the green (dashed-dotted line) one to $\alpha/\beta=0.8$, and the cyan (dashed line) curve represents $\alpha/\beta=0.7$. These last three curves present a strange behavior where there is a change of the relevant projective based Bell inequality ($I_{10}$ to $I_{11}$).}
	\label{fig:DIF_PES}
\end{figure}

To elucidate if PES have advantages over the MES, to observe experimentally the relevance of POVMs in the $I_{CH3}$-inequality violation, we use again the difference between the $I_{CH3}$ maximum obtained with POVMs and with von Neumann measurements. In Fig. \ref{fig:DIF_PES} we observe that the differences $I_{CH3}^{(POVM)} - I_{CH3}^{(PROJ)}$ for $\alpha/\beta<1$ are below the difference $I_{CH3}^{(POVM)} - I_{10}$ for $\alpha/\beta=1$. This implies, that the MES is the optimal state to demonstrate the POVMs relevance in a Bell test of quantum nonlocality based on the $I_{CH3}$-inequality.
\section{Threshold detection efficiencies}
Bell inequalities represent constraint between the probabilities of recording certain events in an experiment. These probabilities can be modified to include the detection efficiency, $\eta$, defined as the ratio between the number of detected events and the number of prepared systems. For this reason, it is always possible to rewrite Bell inequalities in terms of $\eta$ \cite{Eberhard,Garuccio}. The modified Bell inequality can only be violated when the detection efficiency value overcomes a certain critical value, the so-called threshold detection efficiency, $\eta_{crit}$ \cite{Pearle}.

In this last part of our study, we investigate which are the minimum detection efficiencies $\eta_{crit}$ required by the $I_{CH3}$ inequality, and what type of measurements are associated to them, for a loophole-free Bell test. The $\eta_{crit}$ values, considering different degrees of entanglement for the two-qubit state and distinct $c$ values, were calculated using once again the CG method. As previously mentioned, we considered starting point samples in the order of $10^4$ for each simulation, and a numeric precision search of $10^{-6}$.

Following the method of \cite{Eberhard,Garuccio} and considering a symmetric Bell test (i.e., $\eta = \eta_A = \eta_B$), we can rewrite the $I_{CH3}$ inequality as
\begin{equation}
 I_{CH3} = \eta^2 I_{CH3}^{(2)} + \eta(1-\eta)\left(I_{CH3}^{(1A)} + I_{CH3}^{(1B)}\right) + (1-\eta)^2I_{CH3}^{(0)}, \label{ICH3eff}
\end{equation} where $I_{CH3}^{(2)}$, $I_{CH3}^{(1A)}$, $I_{CH3}^{(1B)}$ and $I_{CH3}^{(0)}$ are the values of $I_{CH3}$ when two particles, only Alice's particle, only Bob's particle and no particles are detected, respectively. In the case of $I_{CH3}$, the threshold detection efficiency is given by
\begin{widetext}
\begin{equation}
\eta_{crit} \equiv \frac{c\,p_A(0|0)p_A(0|2)+(1-1/\sqrt{2})p_A(1|2)+(c+1)p_B(0|0)+p_B(0|1)}{c\,[\,p(00|00)+p(00|01)+p(00|10)-p(00|11)]+p(00|20)+p(00|21)+p(10|20)-p(10|21)}.  \label{eq:etacrit}
\end{equation}
\end{widetext}

After performing the $\eta_{crit}$ minimization, we compared the obtained results with the $\eta_{crit}$ values of $I_{CH}$ \cite{ClauserHorne,Eberhard,Larsson2}. When we compare these values for a fixed degree of entanglement, we find that $\eta_{crit}$ of $I_{CH3}$ ($\eta_{crit}~I_{CH3}$) approaches $\eta_{crit}$ of $I_{CH}$ ($\eta_{crit}~I_{CH}$) as $c$ increases. This is due, as previously mentioned, to the fact that $I_{CH}$ becomes dominant as $c$ increases. Surprisingly, however, the measurements that give the lowest value of $\eta_{crit}$ of $I_{CH3}$, according to the numeric minimization done, correspond to \emph{rank-2} POVMs and not to projective measurements, as one could expect at first sight. It is important to note that the program does not force Alice's measurement operators to be three-element POVMs. So, if it were the case that one of the inequalities $I_{00}$, $I_{01}$, $I_{10}$, $I_{11}$, $I_{02}$ or $I_{20}$ gives a threshold efficiency lower than the efficiency required by $I_{CH3}$ while considering POVMs, the program would provide the corresponding projectors. In our results, in none of the simulations (i.e., for all the $c$ values and $\alpha/\beta$ considered) the projective measurements rendered threshold efficiencies lower than POVMs.

\begin{figure}[ht]
	\centering
		\includegraphics[width=0.35\textwidth,angle=-90]{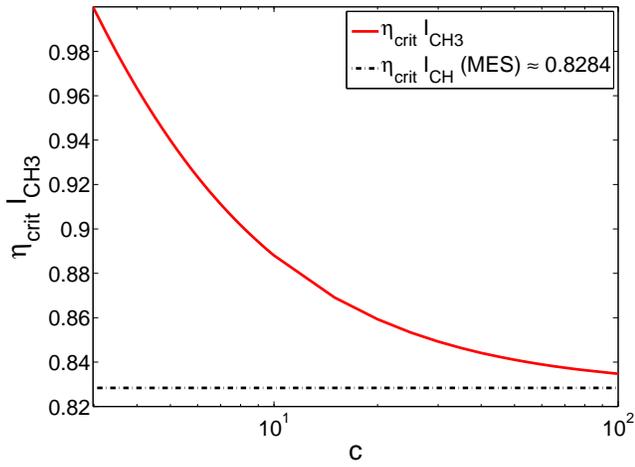}   
	\caption{(Color online) Minimum efficiency required by $I_{CH3}$ as a function of $c$ for the MES for a loophole-free Bell test. We observe the behavior of $\eta_{crit}$ for $I_{CH3}$ as the parameter $c$ increases and compare these values with $\eta_{crit}$ of the $I_{CH}$ inequality.}
	\label{fig:MES_c0_100_EFF}
\end{figure}

Figure~\ref{fig:MES_c0_100_EFF} shows the behavior of $\eta_{crit}~I_{CH3}$ for the maximally entangled state. One can note that the minimal efficiencies required by $I_{CH3}$ tends to the $\eta_{crit}$ value of $I_{CH}$ ($82.8\%$) as $c$ increases. For example, we obtain that $\eta_{crit}~I_{CH3}=0.8348$ when $c=100$. In this case, the minimal efficiency is given by a rank-2 POVM with the following elements
\begin{small}
\begin{eqnarray}
M_0 &=&\left(
\begin{array}{cc}
	0.8009 & -0.0844-0.0262i\\
	-0.0844+0.0262i & 0.0102
\end{array}
\right),  \\
M_1 &=&\left(
\begin{array}{cc}
	0.0703 & -0.0785-0.1673i\\
	-0.0785+0.1673i & 0.4889
\end{array}
\right),
\label{eq:POVMEFFMES}
\end{eqnarray}
\end{small}with $M_2 = \mathds{1} - M_0 - M_1$. These operators are rank-2 within the numeric precision we are considering. The eigenvalues of $M_0$, are $\lambda_{01} = 0.000488$ and $\lambda_{02} =~ 0.810611$, of $M_1$ are $\lambda_{11} = 0.000369$ and $\lambda_{12} = 0.558819$, and of $M_2$ are $\lambda_{21} = 0.000834$ and $\lambda_{22} = 0.628876$.

\begin{figure}[ht]
	\centering
		\includegraphics[width=0.35\textwidth,angle=-90]{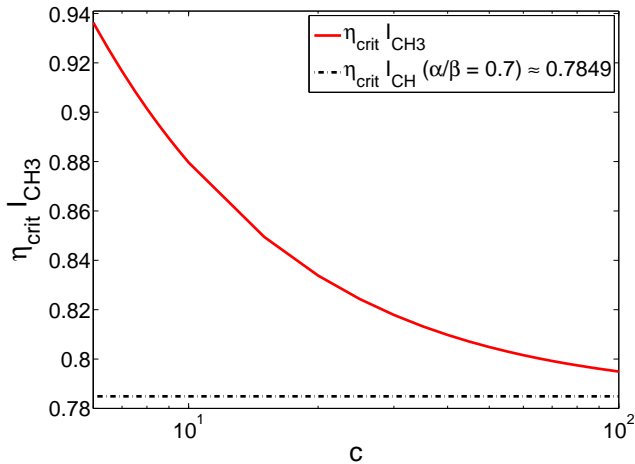}  
	\caption{(Color online) Minimum efficiency required by $I_{CH3}$ as a function of $c$ for a PES, whose $\alpha/\beta = 0.7$. We observe the behavior of $\eta_{crit}$ for $I_{CH3}$ as the parameter $c$ increases and compare these values with $\eta_{crit}$ of the $I_{CH}$ inequality.}
	\label{fig:ab07_c0_100_EFF}
\end{figure}

In Fig. \ref{fig:ab07_c0_100_EFF} we study the behavior of $\eta_{crit}$ considering a partially entangled state for which $\alpha/\beta=0.7$. According to $\eta_{crit}$ of $I_{CH}$, we also notice that $\eta_{crit}~I_{CH3}$ is lower for partially entangled states. Again $\eta_{crit}~I_{CH3}$ decreases and gets closer to $\eta_{crit}~I_{CH}(\alpha/\beta=0.7)\approx 0.7849$ as $c$ increases. For $c=100$, $\eta_{crit}~I_{CH3} = 0.7949$. The corresponding type of measurement is a POVM whose elements are
\begin{small}
\begin{eqnarray}
M_0 &=&\left(
\begin{array}{cc}
	0.7624 & -0.0537-0.1546i\\
	-0.0537+0.1546i & 0.0352
\end{array}
\right),  \\
M_1 &=&\left(
\begin{array}{cc}
	0.2374 & 0.0537-0.1546i\\
	0.0537+0.1546i & 0.1129
\end{array}
\right), \label{eq:POVMEFFPES07}
\end{eqnarray} \end{small}where $M_2 = \mathds{1} - M_0 - M_1$. These operators are also rank-2 within the numeric precision we are considering. The eigenvalues of $M_0$ are $\lambda_{01} = 0.000053$ and $\lambda_{02} =~ 0.797505$, of $M_1$ are $\lambda_{11} = 0.000055$ and $\lambda_{12} = 0.350224$, and of $M_2$ are $\lambda_{21} = 0.000208$ and $\lambda_{22} = 0.851952$.

\begin{figure}[ht]
	\centering
		\includegraphics[width=0.35\textwidth,angle=-90]{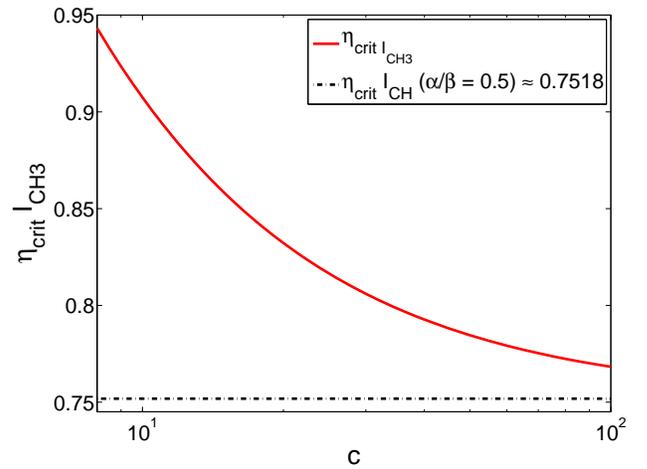}  
	\caption{(Color online) Minimum efficiency required by $I_{CH3}$ as a function of $c$ for PES, whose $\alpha/\beta = 0.5$. We observe the behavior of $\eta_{crit}$ for $I_{CH3}$ as the parameter $c$ increases and compare these values with $\eta_{crit}$ of the $I_{CH}$ inequality.}
	\label{fig:ab05_c0_100_EFF}
\end{figure}

For $\alpha/\beta=0.5$ we have a behavior similar to $\alpha/\beta = 0.7$. In Fig. \ref{fig:ab05_c0_100_EFF} we can see once again that $\eta_{crit}~I_{CH3}$ tend to $\eta_{crit}~I_{CH}(\alpha/\beta=0.5)\approx 0.7518$ as $c$ increases. In $c=100$ we obtain $\eta_{crit}~I_{CH3} = 0.7683$. This result is also given by a rank-2 POVM with the elements being
\begin{small}
\begin{eqnarray}
M_0 &=&\left(
\begin{array}{cc}
	0.7977 & -0.0946+0.0340i\\
	-0.0946-0.0340i & 0.0127
\end{array}
\right), \\
M_1 &=&\left(
\begin{array}{cc}
	0.2020 & 0.0945-0.0341i\\
	0.0945+0.0341i & 0.0500
\end{array}
\right), \label{eq:POVMEFFPES05}
\end{eqnarray}\end{small}with $M_2 = \mathds{1} - M_0 - M_1$. Once again we have rank$-2$ operators within the numeric precision we are considering. The eigenvalues of $M_0$ are $\lambda_{01} = 0.000028$ and $\lambda_{02} =~ 0.810380$, of $M_1$ are $\lambda_{11} = 0.000037$ and $\lambda_{12} = 0.252028$, and of $M_2$ are $\lambda_{21} = 0.000243$ and $\lambda_{22} = 0.937282$.

\begin{figure}[ht]
	\centering
		\includegraphics[width=0.35\textwidth,angle=-90]{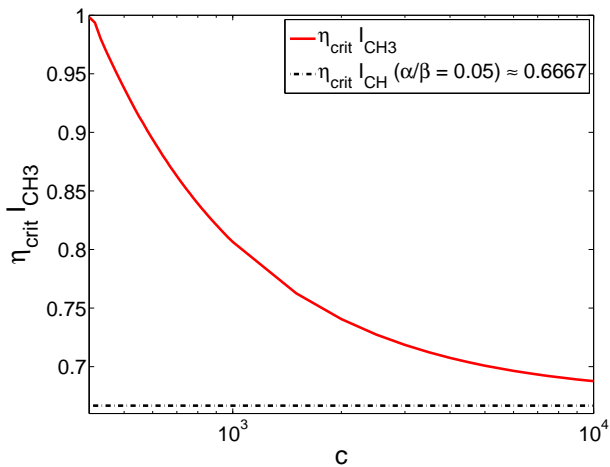}  
	\caption{(Color online) Minimum efficiency required by $I_{CH3}$ in function of $c$ for quasi-product states, where $\alpha/\beta = 0.05$. We observe the behavior of $\eta_{crit}$ for $I_{CH3}$ as the parameter $c$ increases and compare these values with $\eta_{crit}$ of the $I_{CH}$ inequality.}
	\label{fig:ab005_c400_10000_EFF}
\end{figure}

Finally, we analyzed the case of an almost product state with $\alpha/\beta=0.05$. This type of states are important, because in the case of two-qubits, they demand the lowest known detection efficiencies for closing the \emph{detection loophole} in a Bell test \cite{Eberhard,Larsson2}. Fig.~\ref{fig:ab005_c400_10000_EFF} shows the behavior of $\eta_{crit}~I_{CH3}$ for a PES with $\alpha/\beta=0.05$. One can note the same behavior seen for the other analyzed states. We see that as $c$ increases, $\eta_{crit}~I_{CH3}$ values decrease and get closer to the value of $\eta_{crit}~I_{CH}(\alpha/\beta=0.05)\approx 0.6667$. For $c=10000$, we obtained that the minimum efficiency corresponds to $\eta_{crit}~I_{CH3}=0.6876$. This minimal threshold efficiency is given by a clear rank-2 POVM whose elements are
\begin{small}
\begin{eqnarray}
M_0 &=&\left(
\begin{array}{cc}
	0.380029 & -0.000004+0.000002i \\
	-0.000004-0.000002i & 0.002529
\end{array}
\right),  \nonumber \\ \\
M_1 &=&\left(
\begin{array}{cc}
	0.361292 & -0.000004+0.000009i\\
	-0.000004-0.000009i & 0.007412
\end{array}
\right), \nonumber \\ \label{eq:POVMEFFPES005}
\end{eqnarray}
\end{small}where $M_2 = \mathds{1} - M_0 - M_1$. We have that the eigenvalues corresponding to $M_0$ are $\lambda_{01} = 0.002529$ and $\lambda_{02} =~ 0.380029$, those corresponding to $M_1$ are $\lambda_{11} =~ 0.007412$ and $\lambda_{12} = 0.361291$, and to $M_2$ the eigenvalues $\lambda_{21} = 0.258678$ and $\lambda_{22} = 0.990058$.

It is important to notice that even at high $c$ values, when $I_{CH}$ is totally dominant in $I_{CH3}$, the type of measurement associated to $\eta_{crit}~I_{CH3}$ is still a POVM and not a projective measurement. Even though this result may be surprising when first analyzed, it is in agreement with the results shown in Fig. \ref{fig:ICH3_I10_ab05_MES_c3}, Fig. \ref{fig:ICH3_I10_I11_ab05_MES_c5} and Fig. \ref{fig:ICH3_I10_I11_ab05_MES_c10}. In these plots we clearly showed that for higher $c$ values, POVMs give the highest violation for a broader range of states $\alpha/\beta$. Since $\eta_{crit}$ is inversely proportional to the maximal violation of the inequality, it is reasonable to expect that the lowest values of $\eta_{crit}$ are associated to POVMs.

Moreover, it is also worthwhile to mention that as we have used a program that relies in an exhaustive search technique, we can not affirm that the rank-2 POVMs found are the \emph{only} POVMs that give the lowest threshold efficiencies. In accordance with \cite{Gisin2008}, it is reasonable to assume that there should also exist rank-1 POVMs that can attain the same minimal value obtained for the threshold efficiencies.


\section{Conclusion}

In this work we have studied further properties of the Bell-inequality introduced by V\'{e}rtesi and Bene \cite{Vertesi y Bene}, $I_{CH3}$, which they showed to be maximally violated only when more general positive operator valued measures (POVMs) are used instead of the usual von Neumann measurements. For maximally entangled states we performed the maximization of $I_{CH3}$ using a exhaustive numerical search based on the conjugated gradient. Due to the fact that we left in total freedom the choice of Alice's and Bob's measurement operators, we obtained a higher quantum bound of this inequality using POVMs. Once we obtained a higher quantum bound for $I_{CH3}$, we investigated if there is an experimental setup that can be used for observing that POVMs give higher violations in Bell tests based on this inequality. We analyzed the maximum errors supported by the inequality and indicated a source of entangled photons that can be used for the test \cite{Shapiro2004,Kim2006,Smith2012}.

We also studied the maximum violation of the $I_{CH3}$-inequality while considering different degrees of entanglement for a two-qubit state, and different values of the $c$ parameter involved in the inequality. For partially entangled states, we found that as the $c$ value increases, POVMs become relevant (for maximizing the violation of this inequality) for a broader range of entangled states. From this study we obtained that the optimal state for demonstrating the relevance of POVMs in a Bell test based on $I_{CH3}$ is the MES.

Finally, we investigated which are the threshold efficiencies ($\eta_{crit}$) required by this inequality and which are the type of measurements associated to them. We obtained that the values of $\eta_{crit}~I_{CH3}$  are very close to the minimal detection efficiencies known to date for the case of two-qubits, $\eta_{crit}~I_{CH}$. For a fixed degree of entanglement, we find that $\eta_{crit}$ of $I_{CH3}$ tends to $\eta_{crit}$ of $I_{CH}$ as $c$ increases. This is due to the fact that $I_{CH}$ becomes dominant in $I_{CH3}$ as $c$ increases. However, the measurements that give the lowest value of $\eta_{crit}$ of $I_{CH3}$, according to the numeric minimization done, correspond to \emph{rank-2 POVMs} and not to projective measurements. Therefore, we have that POVMs could also be relevant for the recent effort of performing a loophole-free Bell tests.

\begin{acknowledgments}
We gratefully acknowledge T. V\'{e}rtesi for his comments and careful reading of our manuscript. This work was supported by Grants FONDECYT 1120067, Milenio~P10-030-F and PIA-CONICYT PFB0824. J. F. B. and E. S. G. acknowledge the financial support of CONICYT.
\end{acknowledgments}

\end{document}